\documentclass[
superscriptaddress,
 amsmath,
 amssymb,
 aps,
 a4paper,
 floatfix,
 longbibliography,
 pra
]{revtex4-2}

\usepackage{graphicx}% Include figure files
\usepackage{dcolumn}% Align table columns on decimal point
\usepackage{bm}% bold math
\usepackage{physics}

\usepackage{xcolor}

\usepackage{microtype}

\usepackage[american]{babel}
\usepackage[utf8x]{inputenc}

\usepackage{siunitx} 

\newcommand{\ee}{\text{e}}
\newcommand{\ii}{\text{i}}

\definecolor{cset-aps-blueberry}{RGB}{28,128,158}
\definecolor{cset-aps-blue}{RGB}{46,44,184}
\definecolor{cset-aps-turquoise}{RGB}{0,67,88}
\definecolor{cset-aps-limegreen}{RGB}{190,219,67}
\definecolor{cset-aps-green}{RGB}{31,138,112}
\definecolor{cset-aps-yellow}{RGB}{255,225,25}
\definecolor{cset-aps-orange}{RGB}{253,116,0}
\definecolor{cset-aps-red}{RGB}{219,0,43}
\definecolor{myred}{RGB}{255,0,20}

\usepackage{hyperref}
\hypersetup{%
    colorlinks=true,
    linkcolor={cset-aps-red},
    linkbordercolor={cset-aps-red},
    filecolor={cset-aps-orange},
    filebordercolor={cset-aps-orange},
    citecolor={cset-aps-blue},
    citebordercolor={cset-aps-blue},
    urlcolor={cset-aps-blueberry},
    urlbordercolor={cset-aps-blueberry},
    menucolor={cset-aps-limegreen},
    menubordercolor={cset-aps-limegreen},
    breaklinks=true,
    pdfborderstyle={/S/U/W 2},
    pdfpagemode=UseOutlines,
    pdfstartpage={1},
}

\usepackage{cleveref}

\begin{document}

\preprint{APS/123-QED}

\title{A different perspective on the Landau-Zener dynamics}% Force line breaks with \\

\author{Eric P. Glasbrenner}
\affiliation{Institut f{\"u}r Quantenphysik and Center for Integrated Quantum
    Science and Technology (IQ\textsuperscript{ST}), Universit{\"a}t Ulm, Albert-Einstein-Allee 11, D-89081 Ulm, Germany}
\email{eric.glasbrenner@uni-ulm.de}
\author{Yannik Gerdes}%
\affiliation{Institut f{\"u}r Quantenphysik and Center for Integrated Quantum
    Science and Technology (IQ\textsuperscript{ST}), Universit{\"a}t Ulm, Albert-Einstein-Allee 11, D-89081 Ulm, Germany}
\email{yannik.gerdes@uni-ulm.de}
\author{Sándor Varró}
\affiliation{ELI-ALPS (Attosecond Light Pulse Source) Research Institute, ELI-HU Ltd., 6728 Szeged, Hungary}%
\author{Wolfgang P. Schleich}
\affiliation{Institut f{\"u}r Quantenphysik and Center for Integrated Quantum
    Science and Technology (IQ\textsuperscript{ST}), Universit{\"a}t Ulm, Albert-Einstein-Allee 11, D-89081 Ulm, Germany}%
\affiliation{Institute for Quantum Science and Engineering (IQSE), and Texas A\&M AgriLife Research and Hagler Institute for Advanced Study, Texas A\&M University, College Station, TX 77843-4242, USA}

\date{\today}% It is always \today, today,

\begin{abstract}
We present two different approaches towards the Landau-Zener problem: (i) The Markov approximation in the integro-differential equation for one of the two probability amplitudes, and (ii) an amplitude-and-phase analysis of the linear second order differential equation for same probability amplitude. Our treatment shows that the Markov approximation neglects the non-linearity of the equation but still provides us with the exact asymptotic result.
\end{abstract}

\maketitle

\section{Introduction}

‘I’ll give you a definite maybe’. This quote by the well-known Polish-born American film producer Samuel Goldwyn, may well be interpreted as a one-line summary of the superposition principle of quantum mechanics. Nowhere clearer do we see quantum interference at work than in the Landau-Zener effect \cite{Landau1932a, Landau1932b, Zener1932, Majorana1932, Stueckelberg1932, Kofman2023},  with applications ranging from nuclear fission \cite{Hill1953} via molecular conical intersections \cite{herzberg1939} to cold atoms in optical lattices \cite{Liu2002, Christiani2002, Konotop2005, Zenesini2009}, matter wave interferometry \cite{SHEVCHENKO2010, ivakhnenko2023, konrad2024} and driven quantum systems \cite{Wubs2006, kofman2024}.
\par
In a recent article \cite{Glasbrenner2023}, we have obtained two ‘one-line derivations’ of the exact Landau-Zener probability amplitude. The first one was based on the Markov approximation \cite{scullyzubairy1997}, and the other on neglecting a second derivative and a deeper understanding of the logarithmic phase singularity. In the present article, we provide a different perspective on the Landau-Zener formula by first expanding on our approach based on the Markov approximation, and then comparing and contrasting it to an amplitude-and-phase approach. 
\par 
It is a great honor and pleasure for us to dedicate our contribution to this volume to Professor Victor Dodonov on the occasion of his 75th birthday. Our paths have frequently crossed, and we have learned so much from his asymptotic analysis of the oscillatory photon statistics of a squeezed state, his group theoretical approach towards quantum optics, and his deep insights into the dynamical Casimir effect.
\par
Since at the very heart of Victor’s research is asymptotology, we have chosen as the topic for our birthday essay the Landau-Zener effect and hope that it will find his interest. Happy birthday Victor, and many more healthy years in science!
\par
Our article is organised as follows:
In section \ref{section:2}, we briefly formulate the problem of Landau-Zener transitions and present the two coupled differential equations of first order for the two probability amplitudes in the interaction picture. Throughout this article we focus on the dynamics of a single probability amplitude.
\par
We dedicate section \ref{sec:markov:approximation} to a discussion of the approximate but analytic solution of the integro-differential equation for one of the probability amplitudes. This approach provides us with the exact expression for the transition probability amplitude. Moreover, we derive approximate but analytic expressions for the Markov solution, which we compare and contrast to the exact numerical ones. In particular, we establish a relation which connects the integrand of the Markov solution at negative times with positive times. This analysis brings out most clearly the Stueckelberg oscillations. 
\par
In section \ref{sec:amplitude:phase:approach}, we pursue a different approach and first derive from the integro-differential equation a linear differential equation of second order. We then obtain two coupled differential equations of second order for the absolute value and the phase velocity of the probability amplitude. Although these equations are rather complicated, we can solve them after a linearization. In this way, we make contact with the expressions obtained with the help of the Markov solution. 
\par
We conclude by summarizing our results in section \ref{section:conclusion:outlook}, and provide a brief outlook. 

\section{Landau-Zener transitions}
\label{section:2}
In the present section we summarize the Landau-Zener problem and introduce dimensional variables. We then present the coupled differential equations for the two probability amplitudes in the Schrödinger as well as in the interaction picture. We conclude by presenting the time dependence of one of them as a trajectory in the complex plane based on a numerical integration of the coupled differential equations.

\subsection{Formulation of the problem}
In its most elementary form the Landau-Zener problem is given by the Schrödinger equation
\begin{align}
    \ii\hbar \frac{\dd}{\dd t}\begin{pmatrix}
        \Tilde{a}(t) \\
        \Tilde{b}(t)
    \end{pmatrix} = H(t)
    \begin{pmatrix}
        \Tilde{a}(t) \\
        \Tilde{b}(t)
    \end{pmatrix} 
    \label{eq:Schrödinger:equation}
\end{align}
governed by the time-dependent Hamiltonian
\begin{align}
    \hat{H}(t) = \hbar
    \begin{pmatrix}
    -\alpha t & \Omega \\
    \Omega & \alpha t
    \end{pmatrix}.
    \label{eq:time:dependent:hamiltonian}
\end{align}
Here, $\alpha$ and $\Omega$ denote the chirp rate and the coupling constant.
\par
With the dimensionless time $\tau \equiv \Omega\cdot t$, the probability amplitudes $a=a(\tau)$ and $b=b(\tau)$ follow from the set of two coupled differential equations
\begin{align}
    \ii\dot{\Tilde{a}}(\tau) &= -\epsilon\tau\Tilde{a}(\tau) + \Tilde{b}(\tau)\\
    \ii\dot{\Tilde{b}}(\tau) &= \Tilde{a}(\tau) + \epsilon\tau \Tilde{b}(\tau)
\end{align}
where the dot denotes the differentiation with respect to $\tau$. Here we have introduced the new parameter $\epsilon \equiv \alpha/\Omega^{2}$ as ratio between the chirp $\alpha$ and the square of the coupling constant $\Omega$.
\par
It is useful to transform from the Schrödinger picture defined by \cref{eq:Schrödinger:equation} into an interaction picture using the definitions
\begin{align}
    \Tilde{a}(\tau) &\equiv \exp\left(\frac{\ii}{2}\epsilon\tau^{2}\right)a(\tau)\\
    \Tilde{b}(\tau) &\equiv \exp\left(-\frac{\ii}{2}\epsilon\tau^{2}\right)b(\tau)
\end{align}
which yields the set of coupled differential equations
\begin{align}
    \label{eq:diff:equation:a}
    \ii\dot{a}(\tau) &= \ee^{-\ii\epsilon\tau^{2}}b(\tau)\\
    \ii\dot{b}(\tau) &= \ee^{\ii\epsilon\tau^{2}}a(\tau).
    \label{eq:diff:equation:b}
\end{align}

Throughout this article, we consider solutions of these equations for the initial condition 
\begin{align}
    a(\tau\rightarrow -\infty) = 1
    \label{eq:initial:condition:a}
\end{align}
which with the help of the normalization relation 
\begin{align}
    \abs{a(\tau)}^{2} + \abs{b(\tau)}^{2} = 1
    \label{eq:normalization:relation}
\end{align}
translates into the initial condition 
\begin{align}
    b(\tau\rightarrow -\infty) = 0
    \label{eq:initial:condition:b}
\end{align}
for the probability amplitude $b$.

\subsection{Trajectory of probability amplitude in the complex plane}
\begin{figure}[ht]
	\includegraphics[width=9.5cm]{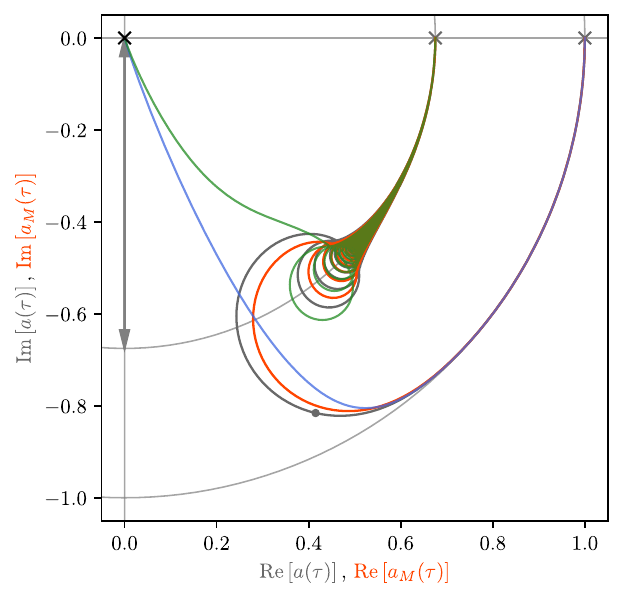}
	\centering
	\caption[]{Probability amplitudes of a Landau-Zener transition with (red line) and without (dark grey line, numerical solution) the Markov approximation represented in the complex plane for the parameters $\alpha = 0.25$ and $\Omega = 0.25$, that is $\epsilon = 4.0$. The dark grey dot marks $\tau=0$. Along the trajectory we observe a slight deviation between the numerical and approximated curve especially during the transition from the unit circle to the one with radius (grey arrow) $a(\tau\rightarrow\infty)= \exp(-\pi/(2\epsilon))$. In the limit for $\tau\rightarrow\pm\infty$ both curves agree almost perfectly. The blue and green lines indicate the elementary analytic approximations, \cref{eq:approxmation:for:large:negative:times} and \cref{eq:expression:eta:M:positive:time}, for large negative and positive times. In the neighborhood of the transition, that is $\tau = 0$, the approximation breaks down completely due to the singularity of these approximation of $\eta_{M}$ at $\tau = 0$.}
	\label{fig:1}
\end{figure}

In \cref{fig:1} we depict the time dependence of the probability amplitude $a$ resulting from the numerical integration of the set of equations, \cref{eq:diff:equation:a} and \cref{eq:diff:equation:b} as a trajectory in the complex plane. In this way, we can interpret the motion 
\begin{align}
    a(\tau) \equiv A(\tau)\ee^{\ii\varphi(\tau)}
    \label{eq:abs:a:e:phi}
\end{align}
by the time-dependent distance $A\equiv\abs{a(\tau)}$ from the origin and the phase angle $\varphi\equiv\varphi(\tau)$ with respect to the real axis.
\par
Due to the initial condition, \cref{eq:initial:condition:a}, the trajectory starts at $\tau = -\infty$ from the value one on the real axis and follows for a long time a circle of radius unity. During this motion the phase angle $\varphi$ increases, starting from $\varphi(\tau = -\infty) = 0$. In this domain the phase velocity $\dot{\varphi}$ is negative. 

However, in the neighborhood of $\tau = 0$ there is a transition from the initial circle of radius unity to another one whose radius is determined by the final value 
\begin{align}
    a(\tau\rightarrow\infty) = \exp\left(-\frac{\pi}{2\epsilon}\right)
    \label{eq:landau:zener:formula:first}
\end{align}
of the probability amplitude given by the Landau-Zener result. Around $\tau = 0$ the phase angle $\varphi$ assumes a maximum and then decreases again. Obviously, the phase velocity $\dot{\varphi}$ switches from purely negative values to an oscillation between positive and negative values. 

Indeed, for positive times, the trajectory performs circular motions, which give rise to the Stueckelberg oscillations. In this regime, the absolute value of $a$ as well as the phase $\varphi$ oscillate. However, the center of the circles approaches the final radius and the end point of the motion on the real axis. Here, $a(\tau = \infty)$ is real again as indicated by \cref{eq:landau:zener:formula:first}.

\section{Markov approximation}
\label{sec:markov:approximation}
In this section, we decouple the set of equations, \cref{eq:diff:equation:a} and \cref{eq:diff:equation:b}, and arrive at an integro-differential equation for the probability amplitude $a$. When we solve this equation with the help of the Markov approximation \cite{scullyzubairy1997}, we immediately obtain the well-known Landau-Zener formula, \cref{eq:landau:zener:formula:first}.
\par
Moreover, we find approximate but analytic results in the limit of large positive and large negative times. These expressions allow us to make the connection to the amplitude and phase approach explored in the next section. 

\subsection{Integro-differential equation}
We start the discussion by converting the set of two coupled differential equations of first order for $a$ and $b$ into a integro-differential equation for $a$.
For this purpose, we formally integrate  \cref{eq:diff:equation:b} for $b$ subjected to the initial condition, \cref{eq:initial:condition:b},  and substitute the result into \cref{eq:diff:equation:a} for $a$, leading us to the integro-differential equation
\begin{align}
    \dot{a}(\tau) = -\ee^{-\ii\epsilon\tau^{2}}\int_{-\infty}^{\tau}\dd\tau^{\prime}\ee^{\ii\epsilon\tau^{\prime 2}}a(\tau^{\prime})
    \label{eq:integro:differential:equation:a}
\end{align}
with the initial condition, \cref{eq:initial:condition:a}.

On first sight this equation looks rather complicated and no analytic insight offers itself. However, the Markov approximation \cite{scullyzubairy1997} 
\begin{align}
    a(\tau^{\prime}) \approx a(\tau),
    \label{eq:markov:approximation}
\end{align}
which assumes that the main contribution of the probability amplitude $a$ to the integral in  \cref{eq:integro:differential:equation:a} arises from the upper limit, allows us to rederive in essentially one line the Landau-Zener formula, \cref{eq:landau:zener:formula:first}. At this point, we do not discuss the validity of this approximation, but rather show that in the asymptotic limit of $\tau\rightarrow\infty$ we arrive at the exact result.

Indeed, the Markov approximation allows us to factor $a$  out of the integral and arrive at the equation
\begin{align}
    \dot{a}_{M}(\tau) = -\eta_{M}(\tau)a_{M}(\tau)
    \label{eq:differential:equation:markov:eta:a}
\end{align}
where we have introduced the Markov rate function 
\begin{align}
     \eta_{M}(\tau) \equiv \ee^{-\ii\epsilon\tau^{2}}\int_{-\infty}^{\tau}\dd\tau^{\prime}\ee^{\ii\epsilon\tau^{\prime 2}}.
     \label{eq:definition:eta:M}
\end{align}
Here, we have included a subscript $M$ to bring out most clearly that the differential equation,  \cref{eq:differential:equation:markov:eta:a}, is an approximation of the exact integro-differential equation,  \cref{eq:integro:differential:equation:a}, based on the Markov approximation, \cref{eq:markov:approximation}.

It is straightforward to integrate \cref{eq:differential:equation:markov:eta:a}, and we find the Markov solution
\begin{align}
    a_{M}(\tau) = \exp\left(-\int_{-\infty}^{\tau}\dd\tau^{\prime}\eta_{M}(\tau^{\prime})\right)
    \label{eq:Markov:solution}
\end{align}
which satisfies the initial condition, \cref{eq:initial:condition:a}. 
\par
We conclude by representing the Markov solution, \cref{eq:Markov:solution}, in amplitude and phase, that is
\begin{align}
    a_{M}(\tau) = A_{M}(\tau)\,\ee^{\ii\varphi_{M}(\tau)}
\end{align}
where 
\begin{align}
    A_{M}(\tau) \equiv \exp\left(-\int_{-\infty}^{\tau}\dd\tau^{\prime}\Re\left[\eta_{M}(\tau^{\prime})\right]\right)
    \label{eq:amplitude:A:M}
\end{align}
and 
\begin{align}
    \varphi_{M}(\tau) =  \int_{-\infty}^{\tau}\dd\tau^{\prime} \dot{\varphi}_{M}(\tau^{\prime}) = -\int_{-\infty}^{\tau}\dd\tau^{\prime} \Im\left[\eta_{M}(\tau^{\prime})\right].
    \label{eq:phase:varphi:M}
\end{align}
Here, we have chosen $\varphi(\tau\rightarrow -\infty) = 0$.
\par
Hence, the real part of $\eta_{M}$ determines after integration, the amplitude $A_{M}$ of $a_{M}$. In particular, when  $\Re[\eta_{M}]$ is positive, the amplitude decays as a function of time. 
\par
The phase $\varphi_{M}$ of $a_{M}$ follows from the integral of the phase velocity $\dot{\varphi}_{M}$ which is the negative of the imaginary part of $\eta_{M}$. Hence, whenever $\Im[\eta_{M}]$ is positive,  $\dot{\varphi}_{M}$ is negative, indicating a motion in the clockwise direction.

\subsection{Exact Landau-Zener formula}
The preceding analysis shows that in the Markov approximation, \cref{eq:markov:approximation}, the complete information about the Landau-Zener transition is encoded in the Markov rate function $\eta_{M}$ given by \cref{eq:definition:eta:M}. For this reason, we will discuss the behavior of $\eta_{M}$ in section \ref{sec:Markov:Approx:analytic:expression} in more detail. However,  
we first demonstrate that despite the Markov approximation, the asymptotic limit of $a_{M}(\tau\rightarrow\infty)$ is identical to the exact result, \cref{eq:landau:zener:formula:first}. 
\par
Indeed, in the asymptotic limit the Markov solution, \cref{eq:Markov:solution} reads
\begin{align}
    a_{M}(\tau\rightarrow\infty) = \ee^{-I}
\end{align}
with the integral
\begin{align}
    I \equiv \int_{-\infty}^{\infty}\dd\tau^{\prime}\ee^{-\ii\epsilon\tau^{\prime 2}}\int_{-\infty}^{\tau^{\prime}}\dd\tau^{\prime\prime}\ee^{\ii\epsilon\tau^{\prime\prime 2}}.
\end{align}
\par
In order to evaluate $I$, we cast it in the form
\begin{align}
    I = \int_{-\infty}^{\infty}\dd\tau\ee^{-\ii\epsilon\tau^{2}}\int_{-\infty}^{\infty}\dd\tau^{\prime}\ee^{\ii\epsilon\tau^{\prime 2}} - \int_{-\infty}^{\infty}\dd\tau\ee^{-\ii\epsilon\tau^{2}}\int_{\tau}^{\infty}\dd\tau^{\prime}\ee^{\ii\epsilon\tau^{\prime 2}}
\end{align}
which with the new integration variables $\Bar{\tau}\equiv-\tau$ and $\Bar{\tau}^{\prime} \equiv -\tau^{\prime}$ in the second integral, and the integral relation
\begin{align}
    \int_{-\infty}^{\infty}\dd\tau\,\ee^{\pm\ii\epsilon\tau^{2}} = \sqrt{\frac{(\pm\ii)\pi}{\epsilon}},
    \label{eq:integral:relation}
\end{align}
reduces to the identity
\begin{align}
    I = \frac{\pi}{\epsilon} - I,
\end{align}
or 
\begin{align}
    I = \frac{\pi}{2\epsilon},
    \label{eq:for:I}
\end{align}
and thus yields the exact Landau-Zener formula, \cref{eq:landau:zener:formula:first}.

\subsection{Properties of the Markov rate function}
\label{sec:Markov:Approx:analytic:expression}
In the preceding subsection, we have obtained an explicit expression, \cref{eq:Markov:solution}, for the Markov solution in terms of the time integral of the Markov rate function $\eta_{M}$ defined by \cref{eq:definition:eta:M}. In this section, we first derive a linear differential equation of first order for $\eta_{M}$, and then obtain approximate formulae for large negative and large positive times. Our analysis brings out most clearly that the Stueckelberg oscillations introduce an asymmetry between these time domains. 

\subsubsection{Differential equation}
We start our analysis of $\eta_{M}$ by deriving a differential equation for $\eta_{M}$ which will become important in the amplitude-and-phase approach pursued in \ref{sec:amplitude:phase:approach}. For this purpose, we differentiate the definition, \cref{eq:definition:eta:M}, of $\eta_{M}$ and arrive immediately at the linear differential equation
\begin{align}
    \dot{\eta}_{M} + 2\ii\epsilon\tau\eta_{M} - 1 = 0
    \label{eq:differential:equation:eta:M}
\end{align}
of first order.
\par
The definition, \cref{eq:definition:eta:M}, of $\eta_{M}$ also determines the initial condition
\begin{align}
    \eta_{M}(\tau\rightarrow -\infty) = 0.
    \label{eq:initial:condition:eta:M}
\end{align}

\subsubsection{Large negative times}
Next, we start from the definition, \cref{eq:definition:eta:M}, of $\eta_{M}$ and consider large negative times $\tau = -\abs{\tau}$ which yields 
\begin{align}
    \eta_{M}(-\abs{\tau}) = \ee^{-\ii\epsilon\tau^{2}}\int_{\abs{\tau}}^{\infty}\dd\Bar{\tau}\,\ee^{\ii\epsilon\Bar{\tau}^{2}} = \ee^{-\ii\epsilon\tau^{2}} \mathcal{F}(\abs{\tau}).
    \label{eq:eta:large:negative:times}
\end{align}
Here, we have first introduced the integration variable $\Bar{\tau} = - \tau^{\prime}$ and then the definition 
\begin{align}
    \mathcal{F}(\tau) \equiv \int_{\tau}^{\infty}\dd\Bar{\tau}\,\ee^{\ii\epsilon\Bar{\tau}^{2}}
\end{align}
of the Fresnel integral \cite{born2019}.
\par
From the asymptotic expansion \cite{abramowitz1965}
\begin{align}
    \mathcal{F}(\tau) = \left[ -\frac{1}{2 \ii \epsilon \tau} + \frac{1}{4 \epsilon^2 \tau^3} + \mathcal{O}\left( \frac{1}{\tau^5} \right) \right] \ee^{\ii \epsilon \tau^2}
\end{align}
of $\mathcal{F}$ we find 
\begin{align}
    \eta_{M}(-\abs{\tau}) \approx \frac{1}{4\epsilon^{2}\abs{\tau}^{3}}+\ii\frac{1}{2\epsilon\abs{\tau}}.
    \label{eq:approxmation:for:large:negative:times}
\end{align}
Therefore, for large negative times the real part of $\eta_{M}$ is positive and according to \cref{eq:amplitude:A:M} the amplitude starts to decrease from unity for decreasing $\tau$. Moreover, the imaginary part of $\eta_{M}$ is also positive and leads, due to \cref{eq:phase:varphi:M}, to a negative phase velocity. Both results are in complete agreement with the numerical integration of the equations of motion shown in \cref{fig:1}.

\subsubsection{Large positive times}

For large positive times, we first extend the integration to infinity and then subtract the additional term to arrive at the expression 
\begin{align}
    \eta_{M}(\tau) \equiv \ee^{-\ii\epsilon\tau^{2}}\int_{-\infty}^{\infty}\dd\tau^{\prime}\ee^{\ii\epsilon\tau^{\prime 2}}- \ee^{-\ii\epsilon\tau^{2}}\int_{\tau}^{\infty}\dd\tau^{\prime}\ee^{\ii\epsilon\tau^{\prime 2}}
\end{align}
which reduces with the integral relation, \cref{eq:integral:relation}, to 
\begin{align}
    \eta_{M}(\abs{\tau}) = -\eta_{M}(-\abs{\tau}) + \sqrt{\frac{\ii\pi}{\epsilon}} \ee^{-\ii\epsilon\tau^{2}}.
    \label{eq:expression:eta:M:positive:time}
\end{align}
In the last step we have compared the second integral to the one in \cref{eq:eta:large:negative:times}. 
\par 
The relation, \cref{eq:expression:eta:M:positive:time}, connects the two time domains of large negative and large positive times. Indeed, for large positive times $\eta_{M}$ consists of two contributions: (i) A contribution which is identical to negative of $\eta_{M}$ at large negative times, and (ii) an oscillatory term with a quadratic phase and a phase shift of $\pi/4$ moving in the complex plane in the clockwise direction  as indicated by \cref{fig:1}. 

\subsection{Landau-Zener probability emerging from Stueckelberg oscillations}
The oscillatory term of $\eta_{M}$ when integrated over time yields the Stueckelberg oscillations and in the limit of infinite positive time the exact Landau-Zener transition probability, \cref{eq:landau:zener:formula:first}. To bring out this fact most clearly, we integrate, \cref{eq:expression:eta:M:positive:time}, over positive times only which yields
\begin{align}
    \int_{-\infty}^{\infty}\dd\tau\,\eta_{M}(\tau) = \sqrt{\frac{\ii\pi}{\epsilon}} \int_{0}^{\infty}\dd\tau\,\ee^{-\ii\epsilon\tau^{2}}.
\end{align}
Here, we have combined the negative and positive times of $\eta_{M}$ into a single integral.
\par
With the help of the integral relation, \cref{eq:integral:relation}, we arrive at 
\begin{align}
    \int_{-\infty}^{\infty}\dd\tau\,\eta_{M}(\tau) = \sqrt{\frac{\ii\pi}{\epsilon}} \cdot\frac{1}{2} \sqrt{\frac{(-\ii)\pi}{\epsilon}} = \frac{\pi}{2\epsilon}
\end{align}
which is real since the imaginary units compensate each other.
\par
As result, the imaginary part of the Markov rate function $\eta_{M}$ must cancel out when integrated over the whole time domain, that is, 
\begin{align}
    \int_{-\infty}^{\infty}\dd\tau\,\Im\left[\eta_{M}\right] = -\int_{-\infty}^{\infty}\dd\tau\,\dot{\varphi}_{M} = 0.
    \label{eq:integral:over:varphi:dot:whole:time}
\end{align}
Moreover, the corresponding time integral of the oscillations yields a contribution which apart from a factor of $1/2$ is identical to the amplitude of the oscillations given by $\sqrt{\pi/\epsilon}$. The factor $1/2$ is a consequence of the fact that the oscillations only occur for positive times.
\par
We will return to these features in the amplitude-and-phase approach discussed in section \ref{sec:amplitude:phase:approach}.

\subsection{Comparison of Markov solution to the exact probability amplitude}

In order to gain insight into the validity of the Markov approximation, we compare and contrast in \cref{fig:1} the exact expression for the probability amplitude $a$, resulting from the numerical integration of the set of equations, \cref{eq:diff:equation:a} and  \cref{eq:diff:equation:b}, to the Markov solution $a_{M}$ given by \cref{eq:Markov:solution}.
\par
Due to the initial condition, \cref{eq:initial:condition:a}, both trajectories start at $\tau = -\infty$ from the value one on the real axis and are almost indistinguishable following a circle of radius of unity. However, in the neighborhood of $\tau = 0$ where the transition between the two circles takes place, the two trajectories slightly deviate from each other. For positive times both trajectories approach each other again and the final value of $a_{M}(\tau = \infty)$ agrees with $a(\tau = \infty)$. 

\section{Amplitude and phase approach}
\label{sec:amplitude:phase:approach}
In \cref{fig:1}, we have depicted the time dependence of the probability amplitude $a$ as a trajectory in the complex plane. This representation leads to an intuitive decomposition of $a$ into its amplitude $A$ and phase $\varphi$.
\par
In this section, we derive the corresponding coupled equations of motion for $A$ and $\varphi$. It is remarkable that one of them can be integrated immediately, and when we substitute its formal solution into the remaining equation of the amplitude-and-phase approach, we arrive at a non-linear differential equation for either $\dot{\varphi}$ or $A$. We conclude this discussion by solving the linearized equation for the phase velocity and make contact with the Markov solution, \cref{eq:Markov:solution}.

\subsection{Linear second-order differential equation}
We start by differentiating the integro-differential equation, \cref{eq:integro:differential:equation:a}, one more time which leads us immediately to the linear second-order differential equation
\begin{align}
    \ddot{a} + 2\ii\epsilon\tau\dot{a} + a = 0,
    \label{eq:differential:equation:second:order:a}
\end{align}
subjected to the initial conditions, \cref{eq:initial:condition:a} and
\begin{align}
    \dot{a}(\tau\rightarrow -\infty) = 0
    \label{eq:initial:condition:a:dot}
\end{align}
which follows from \cref{eq:diff:equation:a} in combination with the initial condition, \cref{eq:initial:condition:b} for $b$. 
\par
The solutions of \cref{eq:differential:equation:second:order:a} are well known to be the parabolic cylinder functions \cite{abramowitz1965}. Indeed, they are at the very heart of the derivation of the exact Landau-Zener formula, \cref{eq:landau:zener:formula:first}. Two linearly independent solutions are needed to satisfy the initial conditions, \cref{eq:initial:condition:a} and \cref{eq:initial:condition:a:dot}. In the limit of $\tau\rightarrow\infty$, this combination provides us with the exact result, \cref{eq:landau:zener:formula:first}.
\par
In Ref. \cite{Glasbrenner2023} we have used \cref{eq:differential:equation:second:order:a} to obtain another one-line derivation of the exact Landau-Zener expression, \cref{eq:landau:zener:formula:first}. For this purpose, we have neglected in \cref{eq:differential:equation:second:order:a}, the second derivative of $a$, leading us to the approximate differential equation
\begin{align}
    2\ii\epsilon\tau\dot{a} = -a.
    \label{eq:differential:equation:without:a:dot:dot}
\end{align}
\par
Since at $\tau=0$ the prefactor of $\dot{a}$ vanishes we face a logarithmic phase singularity, in complete analogy to the one of an energy eigenstate of an inverted harmonic oscillator \cite{Heim2013} when expressed in terms of quadrature variables \cite{Ullinger2022}. In both cases we arrive in a natural way at the exponential function of \cref{eq:landau:zener:formula:first}.
\par
This connection to the inverted harmonic oscillator stands out most clearly when we take the square \cite{Varro2024} of the Hamiltonian given by \ref{eq:time:dependent:hamiltonian}. In this way, we arrive \cite{Varro2024} at the energy eigenvalue equation of two inverted harmonic oscillators with complex eigenvalues.
\par 
Since these aspects of the Landau-Zener problem are not at the center of the present article, we postpone their discussion to a future publication. Nevertheless, we emphasize already now that, \cref{eq:differential:equation:without:a:dot:dot} also provides us with an intriguing connection to Hawking radiation \cite{Ullinger2022}.

\subsection{Equations of motion}
In order to solve the linear second-order differential equation, \cref{eq:differential:equation:second:order:a}, and establish the connection with the representation, \cref{eq:abs:a:e:phi}, of $a$ in the complex plane, we make the ansatz 
\begin{align}
    a(\tau) \equiv A(\tau) \ee^{\ii \varphi(\tau)}
    \label{eq:amplitude:and:phase:ansatz:1}
\end{align}
where $A(\tau) \equiv \abs{a(\tau)}$.
\par
With the help of the identities 
\begin{align}
    \dot{a} = \left( \dot{A} + \ii A \dot{\varphi}\right) \ee^{\ii \varphi}
    \label{eq:a:dot:ansatz:A:phase}
\end{align}
and 
\begin{align}
    \ddot{a} = \left( \ddot{A} + 2\ii \dot{A} \dot{\varphi} + \ii A \ddot{\varphi} - A \dot{\varphi}^2  \right) \ee^{\ii \varphi }
\end{align}
the equation of motion of $a$, \cref{eq:differential:equation:second:order:a}, takes the form
\begin{align}
    \ddot{A} +  \left( -\dot{\varphi}^2 - 2\epsilon \tau \dot{\varphi} + 1 \right)A + \ii\left[ A \ddot{\varphi} + 2 \dot{A}\left( \dot{\varphi} + \epsilon \tau \right) \right] = 0.
    \label{eq:differential:equation:A}
\end{align}
Here, we have already divided by $\exp(\ii\varphi)$.
\par
When we take the real and imaginary parts of \cref{eq:differential:equation:A} we arrive at the two coupled equations 
\begin{align}
    \ddot{A} + \left( -\dot{\varphi}^2 - 2\epsilon \tau \dot{\varphi} + 1 \right) A = 0
    \label{eq:differential:equation:for:amplitude:A}
\end{align}
and
\begin{align}
    A\ddot{\varphi} + 2 \dot{A} \left( \dot{\varphi} + \epsilon \tau \right) = 0
    \label{eq:amplitude:ansatz:A:phase}
\end{align}
for the amplitude $A$ and the phase $\varphi$ of $a$.
\par
It is interesting that these equations are independent of $\varphi$, only derivatives of $\varphi$ enter. This feature is a consequence of the polar decomposition, \cref{eq:amplitude:and:phase:ansatz:1}, of $a$, that is the fact that $\varphi$ enters through an exponential function, and the original differential equation, \cref{eq:differential:equation:second:order:a}, for $a$ is linear.
\par
The initial condition, \cref{eq:initial:condition:a}, translates into 
\begin{align}
    A(\tau\rightarrow-\infty) = 1
    \label{eq:initial:condition:for:A}
\end{align}
and 
\begin{align}
    \varphi(\tau\rightarrow-\infty) = 0.
    \label{eq:initial:condition:for:varphi}
\end{align}
\par
Moreover, the initial condition for $\dot{a}$, \cref{eq:initial:condition:a:dot}, implies with \cref{eq:a:dot:ansatz:A:phase} the relations
\begin{align}
    \dot{A}(\tau\rightarrow-\infty) = 0
    \label{eq:initial:condition:A:dot}
\end{align}
and 
\begin{align}
    \dot{\varphi}(\tau\rightarrow-\infty) = 0.
    \label{eq:initial:condition:phi:dot}
\end{align}
\par
\subsection{Decoupling the equations of motion}
In the preceding section, we have derived the two coupled differential equations, \cref{eq:differential:equation:for:amplitude:A} and \cref{eq:amplitude:ansatz:A:phase}, for the amplitude $A$ and the phase velocity $\dot{\varphi}$. 
In this section, we decouple them using two different approaches.
\par
For this purpose, we note that \cref{eq:amplitude:ansatz:A:phase} is either a homogeneous differential equation of first order for $A$ in terms of $\dot{\varphi}$, $\ddot{\varphi}$ and $\tau$,
or an inhomogeneous differential equation of first order for $\dot{\varphi}$ in terms of $A$, $\dot{A}$, and $\tau$. In the first case, we obtain with the help of the formal solution
$A = A\left[\dot{\varphi}, \ddot{\varphi}, \tau\right]$ from \cref{eq:differential:equation:for:amplitude:A} a differential equation of second order for $\dot{\varphi}$, whereas in the second case, \cref{eq:differential:equation:for:amplitude:A}, together with the formal solution $\dot{\varphi} = \dot{\varphi}[A, \dot{A}, \tau]$, yields an integro-differential equation for $A$.

\subsubsection{Equation of motion for phase velocity}
In order to decouple the two equations for $A$ and $\dot{\varphi}$, we first note that the ansatz 
\begin{align}
    A(\tau) = \exp[-\frac{1}{2} \int_{-\infty}^{\tau} \dd \tau^{\prime} \frac{\ddot{\varphi}(\tau^{\prime})}{\dot{\varphi}(\tau^{\prime}) + \epsilon \tau^{\prime}}]
    \label{eq:general:solution:for:A}
\end{align}
solves equation \cref{eq:amplitude:ansatz:A:phase} subjected to the initial condition, \cref{eq:initial:condition:for:A}. However, at this moment it is not obvious that the ansatz, \cref{eq:general:solution:for:A}, also satisfies the initial condition, \cref{eq:initial:condition:A:dot}, for $\dot{A}$. 
\par
In order to address this question, we now differentiate $A$ given in the form 
\begin{align}
    A(\tau) = \exp\left[-\int_{-\infty}^{\tau} \dd \tau^{\prime} f(\tau^{\prime})\right]
    \label{eq:amplitude:A:f}
\end{align}
where
\begin{align}
    f(\tau) \equiv \frac{1}{2} \frac{\ddot{\varphi}(\tau)}{\dot{\varphi}(\tau) + \epsilon \tau}
    \label{eq:exponential:term:f}
\end{align}
with respect to time which yields
\begin{align}
    \dot{A} = -A f.
\end{align}
\par
With the initial conditions, \cref{eq:initial:condition:for:A} and \cref{eq:initial:condition:phi:dot}, and assuming that for $\tau\rightarrow -\infty$, the phase acceleration $\ddot{\varphi}$ stays finite, we find due to the linear growth of the denominator
\begin{align}
    f(\tau\rightarrow-\infty) = 0,
\end{align}
and thus the initial condition, \cref{eq:initial:condition:A:dot}.
\par
Next, we arrive with the identities
\begin{align}
    \ddot{A} = Af^{2} - A\dot{f}
\end{align}
as well as 
\begin{align}
    \dot{f} = \frac{1}{2} \frac{(\dot{\varphi} + \epsilon\tau)\dddot{\varphi} - (\ddot{\varphi} + \epsilon)\ddot{\varphi}}{\left(
    \dot{\varphi} + \epsilon \tau\right)^{2}}
    \label{eq:exponential:term:f:dot}
\end{align}
at the relation
\begin{align}
     \ddot{A} = -\frac{1}{2} \frac{(\dot{\varphi} + \epsilon\tau)\dddot{\varphi} - (3/2\ddot{\varphi} + \epsilon)\ddot{\varphi}}{\left(
    \dot{\varphi} + \epsilon \tau\right)^{2}}A.
    \label{eq:identity:for:A:ddot}
\end{align}
\par
When we substitute this formula, \cref{eq:identity:for:A:ddot}, into the equation of motion, \cref{eq:differential:equation:for:amplitude:A}, for $A$, we arrive at the non-linear differential equation 
\begin{align}
    3\ddot{\varphi}^{2} - 2\left(\dot{\varphi} + \epsilon \tau\right)\dddot{\varphi} + 2\epsilon\ddot{\varphi}
    - 4\left(\dot{\varphi} + \epsilon \tau \right)^{4} + 4\left[1 + \left(\epsilon \tau\right)^{2}\right]\left(\dot{\varphi} + \epsilon \tau\right)^{2} = 0
    \label{eq:differential:equation:for:phase}
\end{align}
for the phase $\varphi$ which is of third order. However, since $\varphi$ does not appear explicitly, the differential equation is of second order for the phase velocity $\dot{\varphi}$. 
\par
Hence, in order to solve the linear differential equation for $a$, \cref{eq:differential:equation:second:order:a}, in the representation, \cref{eq:amplitude:and:phase:ansatz:1}, of the amplitude $A$ and the phase $\varphi$, we have to proceed in three steps: (i) First we find the solution of the non-linear differential equation of third order for $\varphi$. (ii) We then find the expressions $\dot{\varphi}$ and $\ddot{\varphi}$ and substitute them into the formula , \cref{eq:exponential:term:f}, for $f$. (iii) In the last step, we integrate $f$ over time, which determines by the relation \cref{eq:amplitude:A:f} the amplitude $A$.

\subsubsection{No pole in the function $f$}
\par
From the definition, \cref{eq:exponential:term:f}, of $f$ we note that the denominator of $f$ might vanish, leading to a singularity in $f$. However, the non-linear differential equation, \cref{eq:differential:equation:for:phase}, for $\dot{\varphi}$ shows that in this case also the numerator given by $\ddot{\varphi}$ vanishes.
\par
Indeed, for times $\tau_{0}$ when $\dot{\varphi}(\tau_{0}) + \epsilon \tau_{0} = 0$ we find from \cref{eq:differential:equation:for:phase} the relation 
\begin{align}
    \left[3\ddot{\varphi}(\tau_{0}) + 2\epsilon\right]\ddot{\varphi}(\tau_{0}) = 0
\end{align}
which yields the two conditions 
\begin{align}
    3\ddot{\varphi}(\tau_{0}) + 2\epsilon = 0
    \label{eq:pole}
\end{align}
or 
\begin{align}
    \ddot{\varphi}(\tau_{0}) = 0.
    \label{eq:ddot:varphi:tau:0}
\end{align}
\par
Equation (\ref{eq:pole}) leads to a non-vanishing second derivative of $\varphi$, that is, 
\begin{align}
    \ddot{\varphi}(\tau_{0}) = -\frac{2}{3}\epsilon
\end{align}
and hence to a pole of $f$.
\begin{figure}[ht]
	\includegraphics[width=13.5cm]{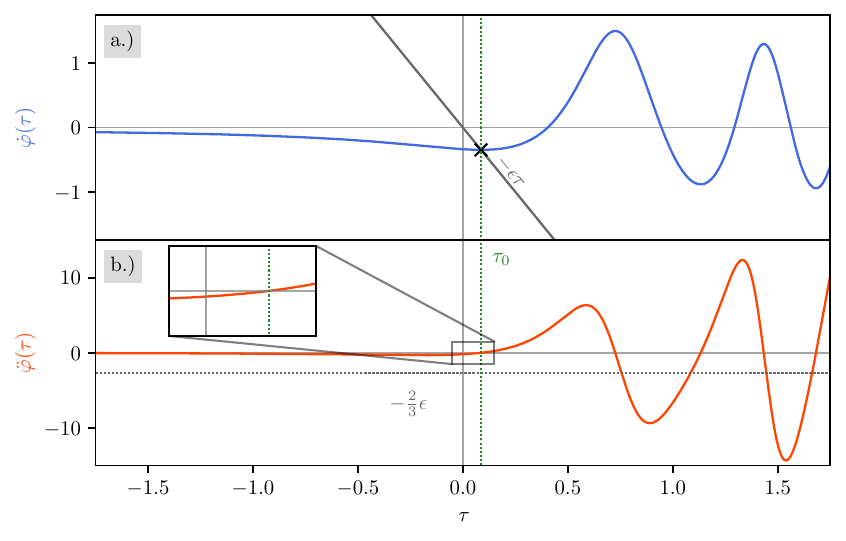}
	\centering
	\caption[No pole in the function $f$]{No pole in the function $f$ governing the amplitude A of the Landau-Zener transition probability amplitude. a.) The denominator of $f$ given by the sum of the phase velocity $\dot{\varphi}$ and $\epsilon\tau$ has a zero at the time $\tau_{0}$ indicated by a vertical green dotted line determined geometrically by the crossing of $\dot{\varphi}$ (blue solid line) and a straight line $-\epsilon\tau$ (dark grey solid line). b.) At $\tau_{0}$, also $\ddot{\varphi}$ (red solid line) vanishes as shown by the inset. As a result, $f$ does not exhibit a singularity at $\tau_{0}$, and in particular, $\ddot{\varphi}$ does not assume the value $-2/3\epsilon$ (dotted grey line) at $\tau_{0}$ nor in its immediate neighborhood but only at later times.}
	\label{fig:2}
\end{figure}
However, as shown in \cref{fig:2} not this case, but \cref{eq:ddot:varphi:tau:0} is realized.
\par
\subsubsection{Equation of motion for amplitude}
We conclude this discussion of the decoupling of the equations for $A$ and $\varphi$ by deriving an equation for $A$ by eliminating $\dot{\varphi}$. For this purpose, we cast \cref{eq:amplitude:ansatz:A:phase} in the form
\begin{align}
        \ddot{\varphi} = -2 \frac{\dot{A}}{A}\dot{\varphi} - 2\frac{\dot{A}}{A}\epsilon \tau
\end{align}
which with the initial condition, \cref{eq:initial:condition:phi:dot}, leads us to the formal solution 
\begin{align}
    \dot{\varphi}(\tau) = -2\epsilon\frac{1}{A^{2}(\tau)}\int_{-\infty}^{\tau}\dd\tau^{\prime}\tau^{\prime}A(\tau^{\prime})\dot{A}(\tau^{\prime})
\end{align}
of $\dot{\varphi}$ in terms of $A$, $\dot{A}$ and $\tau$.
\par
When we substitute this expression into \cref{eq:differential:equation:for:amplitude:A} we find a rather complicated integro-differential equation for $A$ which we do not want to present here. 

\subsection{Linearization of equation for phase velocity}
\label{sec:approximate:solutions}
In the preceding section, we have decoupled the equations, \cref{eq:differential:equation:for:amplitude:A} and \cref{eq:amplitude:ansatz:A:phase}, for the amplitude $A$ and the phase $\varphi$. Unfortunately, the resulting  differential equations are very complicated. 
\par
However, when we neglect in the differential equation, \cref{eq:differential:equation:for:phase}, the non-linear terms we arrive at a linear differential equation which after another approximation is solved by the negative imaginary part of the Markov rate function $\eta_{M}$. Hence, the Markov solution, \cref{eq:Markov:solution}, is at the very heart of the non-linear differential equation, \cref{eq:differential:equation:for:phase}. 

\subsubsection{From a non-linear differential equation to a linear one}
We start by expanding the terms
\begin{align}
   \left(\dot{\varphi} +\epsilon\tau\right)^{4} = \left(\epsilon\tau\right)^{4} + 4 \left(\epsilon\tau\right)^{3}\dot{\varphi} +6 \left(\epsilon\tau\right)^{2}
   \dot{\varphi}^2 + 4\epsilon\tau\dot{\varphi}^3 + \dot{\varphi}^4 
   \label{eq:expand:4}
\end{align}
and 
\begin{align}
    \left(\dot{\varphi} + \epsilon\tau\right)^{2} = \left(\epsilon\tau\right)^{2} + 2\epsilon\tau\dot{\varphi} + \dot{\varphi}^{2}.
    \label{eq:expand:2}
\end{align}
in \cref{eq:differential:equation:for:phase} and neglecting all non-linear terms leading us to the approximate expressions
\begin{align}
    \left(\dot{\varphi} +\epsilon\tau\right)^{4} \approx  \left(\epsilon\tau\right)^{4} + 4 \left(\epsilon\tau\right)^{3}\dot{\varphi}
\end{align}
and 
\begin{align}
    \left(\dot{\varphi} + \epsilon\tau\right)^{2} \approx \left(\epsilon\tau\right)^{2} + 2\epsilon\tau\dot{\varphi}.
\end{align}
\par
Hence, \cref{eq:differential:equation:for:phase} reduces to the inhomogeneous linear differential equation
\begin{align}
    \dddot{\varphi}_{l} -\frac{1}{\tau}\ddot{\varphi}_{l} + \left[\left(2\epsilon\tau\right)^{2} - 4\right]\dot{\varphi}_{l} = 2\epsilon\tau  
    \label{eq:approximate:differential:equation:phi:dot}
\end{align}
of second order for the phase velocity $\dot{\varphi}_{l}$. Here, we have introduced a subscript $l$ to emphasize the fact that we have linearized \cref{eq:differential:equation:for:phase}.
\par
\begin{figure}[ht]
	\includegraphics[width=13.5cm]{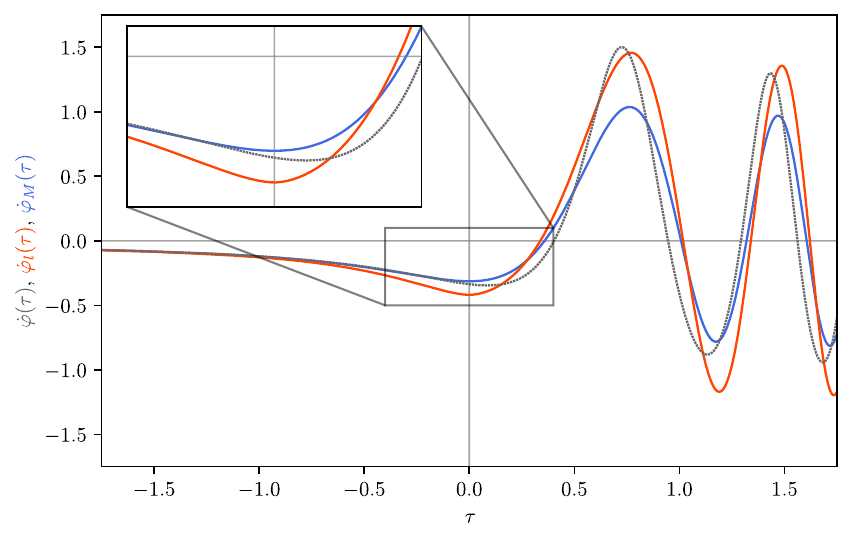}
	\centering
	\caption[]{Influence of the linearization of the non-linear differential equation, \cref{eq:differential:equation:for:phase}, for the phase velocity $\dot{\varphi}$, and the role of the negative shift of the square of the time variable in the linearized equation, \cref{eq:approximate:differential:equation:phi:dot}. The solution of the linear differential equation, \cref{eq:approximate:differential:equation:phi:dot}, (red line), displays a behavior similar to the exact numerical curve (dotted grey line). However, there is a disagreement in the amplitudes and phases. The Markov phase velocity, \cref{eq:varphi:dot:imag:eta:M}, (blue line), which is the exact solution of the linear differential equation, \cref{eq:approximate:differential:equation:phi:dot}, when we neglect the negative shift, that is of \cref{eq:differential:equation:without:minus:four:term}, does not agree in its maxima nor in its phase with the exact solution either. The deviations of the three curves originate from the different behaviors at $\tau=0$ amplified in the inset.}
	\label{fig:4}
\end{figure}

In \cref{fig:4}, we compare and contrast the numerical solution for the phase velocity $\dot{\varphi}_{l}$ of \cref{eq:approximate:differential:equation:phi:dot} represented by the red line with the exact time dependence of $\dot{\varphi}$ given by \cref{eq:differential:equation:for:phase} and shown by the grey dotted line. We note a qualitative but not quantitative agreement. 
\par
\subsubsection{Connection to the Markov solution}
It is interesting that we have already met the exact solution of a differential equation closely related to \cref{eq:approximate:differential:equation:phi:dot}. Indeed, the negative imaginary part of the Markov rate function $\eta_{M}$ solves the differential equation
\begin{align}
    \dddot{\varphi}_{M} - \frac{1}{\tau}\ddot{\varphi}_{M} + \left(2\epsilon\tau\right)^{2}\dot{\varphi}_{M} = 2\epsilon\tau,
    \label{eq:differential:equation:without:minus:four:term}
\end{align}
that is \cref{eq:approximate:differential:equation:phi:dot} in absence of the term $-4\dot{\varphi}_{l}$.
\par
In order to bring this fact out most clearly, we differentiate the differential equation, \cref{eq:differential:equation:eta:M}, of $\eta_{M}$ one more time and arrive at the identity 
\begin{align}
    \ddot{\eta}_{M} + 2\ii\epsilon\eta_{M}+2\ii\epsilon\tau\dot{\eta}_{M} = 0
\end{align}
which with the help of \cref{eq:differential:equation:eta:M} takes the form 
\begin{align}
    \ddot{\eta}_{M} + \frac{1}{\tau}\left(1-\dot{\eta}_{M}\right)+2\ii\epsilon\tau\left(1-2\ii\epsilon\tau\eta_{M}\right) = 0,
\end{align}
or 
\begin{align}
    -\ddot{\eta}_{M} + \frac{1}{\tau}\dot{\eta}_{M} - \left(2\epsilon\tau\right)^{2}\eta_{M} = 2\ii\epsilon\tau + \frac{1}{\tau}.
\end{align}
When we take the negative imaginary part of this equation and make the identification $\Im[\eta_{M}] = -\dot{\varphi}_{M}$, we arrive at \cref{eq:differential:equation:without:minus:four:term}.
\par
Hence, the imaginary part of the Markov rate function $\eta_{M}$ defined by \cref{eq:definition:eta:M}, represents a solution of the differential equation \cref{eq:differential:equation:without:minus:four:term}. Since $\eta_{M}$ is given explicitly, the initial conditions are already fixed. 
\par
The differential equation for $\dot{\varphi}_{M}$, \cref{eq:differential:equation:without:minus:four:term}, is of second order, and thus two initial conditions are necessary. The one for $\dot{\varphi}_{M}$ translates into one for $-\Im[\eta_{M}]$, whereas the one for $\ddot{\varphi}_{M}$ is given by the one for $-\Im[\dot{\eta}_{M}]$. 
\par
As a result, 
\begin{align}
    \dot{\varphi}_{M} = -\Im[\eta_{M}]
    \label{eq:varphi:dot:imag:eta:M}
\end{align}
is a solution of \cref{eq:differential:equation:without:minus:four:term} subjected to the initial condition 
\begin{align}
    \dot{\varphi}_{M}(\tau\rightarrow-\infty) = 0
    \label{eq:initial:condition:varphi:l}
\end{align}
following from the initial condition, \cref{eq:initial:condition:eta:M}, and 
\begin{align}
    \ddot{\varphi}_{M}(\tau\rightarrow-\infty) = 0
    \label{eq:initial:condition:eta:M:dot}
\end{align}
given by the asymptotic expansion, \cref{eq:approxmation:for:large:negative:times} of $\eta_{M}$ together with the differential equation, \cref{eq:differential:equation:eta:M}. 
\par
In \cref{fig:4}, we compare and contrast the Markov phase velocity $\dot{\varphi}_{M}$ given by \cref{eq:varphi:dot:imag:eta:M} and of \cref{eq:Markov:solution} represented by the blue line with the exact time dependence of $\dot{\varphi}$ given by \cref{eq:differential:equation:for:phase} and shown by the grey dotted line. Again we note a qualitative but not quantitative agreement which originates from the time $\tau = 0$ as indicated by the inset. 

\subsection{Decomposition of phase velocity: Anti-symmetric and oscillatory parts}
According to \cref{eq:varphi:dot:imag:eta:M} the imaginary part of the Markov rate function $\eta_{M}$ given by \cref{eq:definition:eta:M} is the exact solution of the differential equation, \cref{eq:differential:equation:without:minus:four:term}, subjected to the initial conditions, \cref{eq:initial:condition:varphi:l} and \cref{eq:initial:condition:eta:M:dot}. However, it is also instructive to find the solutions of the homogeneous and inhomogeneous equation of \cref{eq:differential:equation:without:minus:four:term}. Indeed, they are related to the asymptotic expressions, \cref{eq:approxmation:for:large:negative:times} and \cref{eq:expression:eta:M:positive:time}, of $\eta_{M}$.

\subsubsection{Solutions of homogeneous and inhomogeneous equation}
We first note that the ansatz 
\begin{align}
    \dot{\varphi}_{h}(\tau) \equiv - \mathcal{S} \sin(\beta - \epsilon \tau^2)
    \label{eq:ansatz:phase:velocity:homogenous}
\end{align}
with the two constants $\mathcal{S}$ and $\beta$
solves the homogeneous equation
\begin{align}
    \dddot{\varphi}_{h} -\frac{1}{\tau}\ddot{\varphi}_{h}  + \left(2\epsilon\tau\right)^{2}\dot{\varphi}_{h}  = 0.
    \label{eq:varphi:homogeneous:differential:equation}
\end{align}
\par
Indeed, by differentiation, we find from the identities
\begin{align}
    \ddot{\varphi}_{h}(\tau) = 2 \epsilon \tau \mathcal{S} \cos(\beta - \epsilon \tau^2)
    \label{eq:varphi:dot:dot:homogeneous:solution}
\end{align}
and
\begin{align}
        \dddot{\varphi}_{h}(\tau) = 2 \epsilon \mathcal{S} \cos(\beta - \epsilon \tau^2) + 4\left(\epsilon\tau\right)^{2}\mathcal{S} \sin(\beta - \epsilon \tau^2)
\end{align}
that the ansatz $\dot{\varphi}_{h}$ given by \cref{eq:ansatz:phase:velocity:homogenous} satisfies \cref{eq:varphi:homogeneous:differential:equation}. We emphasize that $\dot{\varphi}_{h}$ given by \cref{eq:ansatz:phase:velocity:homogenous} is an \textit{exact} solution of the \textit{homogeneous} equation, \cref{eq:varphi:homogeneous:differential:equation}
\par
In contrast, the expression 
\begin{align}
    \dot{\varphi}_{i}(\tau) \equiv \frac{1}{2 \epsilon \tau}
    \label{eq:varphi:dot:i}
\end{align}
is a particular solution of the inhomogeneous differential equation, 
\cref{eq:approximate:differential:equation:phi:dot}, only in the limit of large $\abs{\tau}$.
\par
Indeed, by differentiation, we find from the identities
\begin{align}
    \ddot{\varphi}_{i}(\tau) = -\frac{1}{2 \epsilon \tau^2} 
\end{align}
and
\begin{align}
        \dddot{\varphi}_{i}(\tau) = \frac{1}{ \epsilon \tau^3}, 
\end{align}
and by substitution into the differential equation, \cref{eq:differential:equation:without:minus:four:term}, the condition
\begin{align}
    \frac{3}{2\epsilon\tau^{3}} = 0,
\end{align}
which is satisfied for large times.
\par
Hence, the complete asymptotic solution is the sum
\begin{align}
    \dot{\varphi}_{i}(\tau) + \dot{\varphi}_{h}(\tau) = \frac{1}{2 \epsilon \tau} - \mathcal{S} \sin(\beta - \epsilon \tau^2).
    \label{eq:ansatz:large:negative:times:phi:dot}
\end{align}
\par
\subsubsection{Solutions in different time domains}
Since the differential equation, \cref{eq:differential:equation:without:minus:four:term}, is of second order, it contains the two constants $\mathcal{S}$ and $\beta$ of integration. The initial conditions at $\tau\rightarrow -\infty$ determine their values. 
\par
Indeed, for large \textit{negative} times, the solution has to satisfy the initial condition, \cref{eq:initial:condition:phi:dot}, and hence $\mathcal{S}$ has to vanish, leading us to the expression
\begin{align}
    \dot{\varphi}(-\abs{\tau}) \cong \dot{\varphi}_{i}(-\abs{\tau}) \cong -\frac{1}{2\epsilon \abs{\tau}},
    \label{eq:large:negative:times:approximation:1}
\end{align}
in complete agreement with \cref{eq:approxmation:for:large:negative:times} obtained by the Markov approximation. Here we have explicitly taken into account that $\tau$ is negative. We emphasize that in this domain, $\dot{\varphi}$ is solely given by the solution $\dot{\varphi}_{i}$ of the inhomogeneous equation. 
\par
For large \textit{positive} times, the constants of integration $\mathcal{S}$ and $\beta$ have to be determined by extending the solution from large negative times through $\tau = 0$ to large positive times. In this case $\mathcal{S}$ will not vanish, and we end up with \cref{eq:ansatz:large:negative:times:phi:dot}.
\par
When we recall from \cref{eq:large:negative:times:approximation:1} the expression for $\dot{\varphi}$ for large negative times, the solution \cref{eq:ansatz:large:negative:times:phi:dot} for large positive times takes the form 
\begin{align}
    \dot{\varphi}(\tau) = -\dot{\varphi}_{i}(-\abs{\tau}) - \mathcal{S}\sin(\beta - \epsilon \tau^2) 
    \label{eq:ansatz:phi:dot}
\end{align}
which is reminiscent of the imaginary part of \cref{eq:expression:eta:M:positive:time}. 
\par
Hence, to this approximation the phase velocity consists of two parts: (i) An anti-symmetric function given by the inhomogeneous solution $\dot{\varphi}_{i}$, and (ii) an oscillatory function defined by the homogeneous solution $\dot{\varphi}_{h}$ which only is present for positive times.

\subsubsection{Caveats}
In the derivation of the linear differential equations, \cref{eq:approximate:differential:equation:phi:dot} and \cref{eq:differential:equation:without:minus:four:term}, we have neglected all non-linear terms, such as $\dot{\varphi}^{2}$ and higher powers, as well as $\ddot{\varphi}^{2}$ and the product $\dot{\varphi}\dddot{\varphi}$. At this point we have to address the question: Is the resulting solution given by \cref{eq:ansatz:large:negative:times:phi:dot} consistent with this approximation? 
\par
In order to answer the question we consider the solution, \cref{eq:ansatz:phase:velocity:homogenous}, of the homogeneous equation and evaluate the term
\begin{align}
    \ddot{\varphi}^{2}_{h}(\tau) = 4 \epsilon^{2} \tau^{2} \mathcal{S}^{2} \cos^{2}(\beta - \epsilon \tau^2)
\end{align}
which we have neglected with the help of \cref{eq:varphi:dot:dot:homogeneous:solution}. Hence, this contribution displays a secular growth.
\par
Moreover, due to the trigonometric relation
\begin{align}
    \cos^{2}x = \frac{1}{2}\left[1 + \cos\left(2x\right)\right],
\end{align}
the non-linearity creates higher harmonics in the oscillations which are absent in the linearized differential equation, \cref{eq:approximate:differential:equation:phi:dot}, and the Markov rate function, \cref{eq:definition:eta:M}. Although, methods \cite{Carhart1970, Carhart1971, kuehn2015} exist to deal with these two phenomena we do not persue them in this article.
\par
We conclude the discussion of caveats by bringing out one more pecularity. The Markov phase velocity $\dot{\varphi}_{M}$ is an \textit{exact} solution of the \textit{inhomogeneous} equation, \cref{eq:differential:equation:without:minus:four:term}, and only in the limit of large times do we find the decomposition into the oscillatory and anti-symmetric part. Hence, from the point of view of the Markov solution, both the oscillatory and the anti-symmetric part originate from the solution of the \textit{inhomogeneous} equation. 
\par
In contrast, our elementary analysis solving the inhomogeneous differential equation in the two asymptotic time domains identifies the oscillatory part as the solution of the homogeneous equation, and the anti-symmetric part of the inhomogeneous equation.
\par
This on first sight contradictory picture resolves itself when we include the initial conditions. For large negative times the oscillatory part vanishes but only becomes active for positive times.

\subsection{Non-linear corrections of the anti-symmetric part of the phase velocity}
\label{sec:a:glimpse:of:the:non:linearity}
In the preceding section, we have shown that for large \textit{positive} times the negative phase velocity at large \textit{negative} times enters. This result is based on the inhomogeneous solution $\dot{\varphi}_{i}$ defined by \cref{eq:varphi:dot:i}. 
We devote the present section to show that $\dot{\varphi}_{i}$ is only the lowest approximation of a non-linear function given by a square root. In this way, we obtain a relation analogous to \cref{eq:ansatz:phi:dot} but including some aspects of the non-linearity of the equations.
\subsubsection{Sign change of phase velocity}
\par
For this purpose we recall that the numerical integration of the equations of motion leading us to \cref{fig:1} shows that for large negative times the amplitude $A$ does not change significantly, and is roughly given by the initial condition, \cref{eq:initial:condition:for:A}, that is $A(\tau)\approx 1$. For this reason, we can neglect the second derivative of $A$ in \cref{eq:differential:equation:for:amplitude:A}, that is we set $\ddot{A}(\tau) \approx 0 $ which yields
\begin{align}
    \left( -\dot{\varphi}^2 - 2\epsilon \tau \dot{\varphi} + 1 \right) A \approx 0.
\end{align}
\par
Since for large negative times the amplitude is approximately the initial condition and therefore non-zero, this condition translates into the relation
\begin{align}
     -\dot{\varphi}^2 - 2\epsilon \tau \dot{\varphi} + 1 = 0.
     \label{eq:diff:equation:phi:dot}
\end{align}
\par
When we complete the square, the resulting quadratic equation 
\begin{align}
-\left(\dot{\varphi} + \epsilon\tau\right)^{2} + \left(\epsilon\tau\right)^{2} + 1 = 0
\end{align}
yields the obvious solutions
\begin{align}
    \dot{\varphi}(\tau) = -\epsilon \tau \pm \sqrt{\left(\epsilon\tau\right)^{2} + 1}.
    \label{eq:solution:quadratic:equation:phi:dot}
\end{align}
\par
The sign in front of the square root is determined by the initial condition, \cref{eq:initial:condition:phi:dot}. Indeed, for large negative times, \cref{eq:solution:quadratic:equation:phi:dot} reduces to
\begin{align}
    \dot{\varphi}(-\abs{\tau}) = \epsilon \abs{\tau} \left[ 1 \pm \sqrt{1 + \frac{1}{(\epsilon \tau)^2}}\,\right],
\end{align}
which with the well-known asymptotic expansion 
\begin{align}
    \sqrt{1 + x} \approx 1 + \frac{1}{2}x - \frac{1}{8}x^{2}
\end{align}
of a square root gives rise to the expression
\begin{align}
    \dot{\varphi}(-\abs{\tau}) \approx \epsilon \abs{\tau} \left[ 1 \pm \left( 1 + \frac{1}{2(\epsilon \tau)^2} - \frac{1}{8(\epsilon \tau)^4 } \right) \right].
\end{align}
\par
Hence, only the minus sign leads to the correct initial condition, \cref{eq:initial:condition:phi:dot}, with the improved approximation
\begin{align}
    \dot{\varphi}(-\abs{\tau}) \approx  -\frac{1}{2\epsilon \abs{\tau}} +\frac{1}{8(\epsilon\abs{\tau})^3 },
    \label{eq:phi:dot:approx:large:negative:times}
\end{align}
in complete agreement with an naive approach in which we substitute $\dot{\varphi}$ from \cref{eq:large:negative:times:approximation:1} into $\dot{\varphi}^{2}$ in \cref{eq:diff:equation:phi:dot}.
\par
Next we turn to large positive times where \cref{eq:solution:quadratic:equation:phi:dot} takes the form 
\begin{align}
    \dot{\varphi}(\abs{\tau}) = \epsilon \abs{\tau} \left[-1 \pm \sqrt{1 + \frac{1}{(\epsilon \tau)^2}}\,\right]
\end{align}
\par
In order to obtain the first term in \cref{eq:ansatz:phi:dot} we need to take the plus sign, that is 
\begin{align}
    \dot{\varphi}(\abs{\tau}) \approx \frac{1}{2\epsilon \abs{\tau}} -\frac{1}{8(\epsilon\abs{\tau})^3 }.
\end{align}
Hence, in this approximation the phase velocity $\dot{\varphi}$ undergoes at $\tau = 0$ a discontinuous jump from negative to positive values. Moreover, this jump seems to be infinite due to a singularity at $\tau = 0$.

\subsubsection{The jump in the phase velocity: a red herring}
We are now in the position to present an improved approximation for the phase velocity including aspects of the non-linearity of \cref{eq:differential:equation:for:phase}. Since for negative times we have to choose the minus sign in front of the square root in \cref{eq:solution:quadratic:equation:phi:dot}, we find the expression
\begin{align}
    \dot{\varphi}(-\abs{\tau}) = \mathcal{R}(\abs{\tau})
\end{align}
where we have introduced the abbreviation 
\begin{align}
    \mathcal{R}(\abs{\tau}) \equiv \epsilon \abs{\tau} - \sqrt{\left(\epsilon\tau\right)^{2} + 1}.
    \label{eq:definition:R}
\end{align}
\par
Moreover, for positive times we have to select the positive sign in \cref{eq:solution:quadratic:equation:phi:dot}, and the function
\begin{align}
    -\epsilon \abs{\tau} + \sqrt{\left(\epsilon\tau\right)^{2} + 1}  = -\mathcal{R}(\abs{\tau})
\end{align}
is the improved version of the inhomogeneous solution $\dot{\varphi}_{i}$. 
\par
As a result, \cref{eq:ansatz:phi:dot}, including aspects of the non-linearity takes the form 
\begin{align}
   \dot{\varphi}(\tau) = 
   \begin{cases}
     ~~\mathcal{R}(\abs{\tau}) & \text{for}~\tau < 0 \\
     -\mathcal{R}(\abs{\tau}) - \mathcal{S}\sin\left(\beta - \epsilon\tau^{2}\right) & \text{for}~\tau > 0. 
     \label{eq:cases:R:sine}
   \end{cases}
\end{align}
\par
\begin{figure}[ht]
	\includegraphics[width=13.5cm]{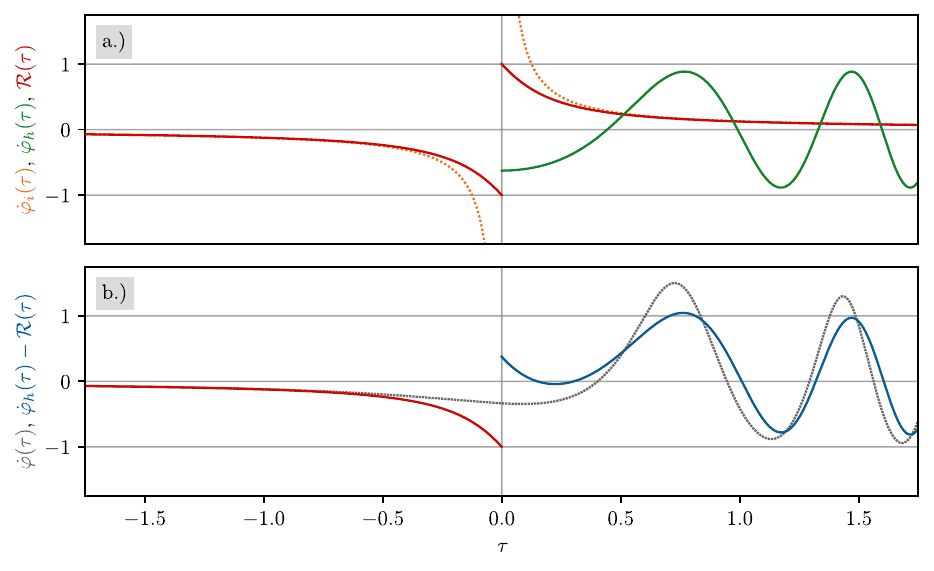}
	\centering
	\caption[]{Cancellation of phase due to the anti-symmetric part of $\dot{\varphi}$ and the phase shift of $\pi/4$ of the Stueckelberg oscillations a.), and comparison of the exact time dependence of the phase velocity $\dot{\varphi}$ with approximations b.). a.) In the most elementary approximation, the phase velocity $\dot{\varphi}$ is the sum of a square root function $\mathcal{R}$, given by \cref{eq:definition:R} (red solid line) which is anti-symmetric with respect to the origin, and an sine function with a quadratic phase and a phase shift of $\pi/4$ (green solid line). This oscillatory function appears only for positive times. The area underneath the curve governing the phase of the transition probability amplitude at $\tau=\infty$ vanishes due to the anti-symmetry of $\mathcal{R}$ and the phase shift of the sine function. b.) The exact expression (gray dotted line) connects in a continuous and differentiable way the two approximations for negative times, given by the negative square root, with the one for positive times given by the sum (blue solid line) of the positive square root and the oscillatory term. For large negative and large positive times the inhomogeneous solutions $\dot{\varphi}_{i}$ given by \cref{eq:varphi:dot:i} are the lowest order approximations of $\mathcal{R}$, as shown in a.) by the orange dotted curve.}
	\label{fig:3}
\end{figure}

This analysis suggests that at $\tau = 0$ there is a \textit{finite} jump in $\dot{\varphi}$ from $-1$ to $1-\mathcal{S}\sin\beta$ as indicated by \cref{fig:3}. Obviously this artifact is a consequence of neglecting the second derivative of $A$. The complete equation, \cref{eq:differential:equation:for:phase}, for $\dot{\varphi}$ connects in a continuous and differentiable way the two time domains as shown in \cref{fig:3}.
\par
Moreover, the square root functions reconfirm the earlier observation that $\dot{\varphi}$ contains an anti-symmetric part, now played by the function $\mathcal{R}$, and an oscillatory part. The comparison of the exact function including the non-linearity of \cref{eq:differential:equation:for:phase} and the elementary sine function displays a deviation as shown in \cref{fig:3}. 

\subsection{Matching the constants of integration}
So far we have not specified the constants of integration $\beta$ and $\mathcal{S}$ which appear in the homogeneous solution $\dot{\varphi}_{h}$ valid for positive times. Unfortunately, it is impossible to obtain $\beta$ and $\mathcal{S}$ since we do not have an initial condition in this time domain. However, we now present a heuristic argument based on matching the asymptotic probability amplitude $a$ in amplitude $A$ and phase $\varphi$ with the Markov solution which is exact.  

\subsubsection{Determination of phase shift}
We start our analysis by considering the phase in the asymptotic limit of an infinite positive time, that is
\begin{align}
    \varphi(\infty) = \int_{-\infty}^{\infty}\dd\tau\,\dot{\varphi}(\tau).
\end{align}
Here, we have taken into account the initial condition, \cref{eq:initial:condition:for:varphi}. 
\par
Next, we recall from \cref{eq:cases:R:sine} that the phase velocity $\dot{\varphi}$ consists of a sum of an anti-symmetric function and an oscillatory contribution that is only active for positive times. Moreover, the anti-symmetric part is the continuous and differentiable extension of the square root functions $\mathcal{R}$, given by \cref{eq:definition:R}.
\par
Hence, the anti-symmetric part of $\dot{\varphi}$ cancels out in this integral and only the oscillatory part $\dot{\varphi}_{h}$ for positive times survives, leading us to the expression
\begin{align}
    \varphi(\infty) = \int_{0}^{\infty}\dd\tau\,\dot{\varphi}_{h}(\tau)
\end{align}
which with the help of \cref{eq:ansatz:phase:velocity:homogenous} takes the form
\begin{align}
     \varphi(\infty) = -\mathcal{S}\mathcal{I}_{s}.
\end{align}
Here, we have introduced the abbreviation 
\begin{align}
    \mathcal{I}_{s} \equiv \int_{0}^{\infty}\dd\tau  \sin(\beta - \epsilon \tau^2) = \Im \left[\ee^{\ii \beta} \int_{0}^{\infty} \dd \tau\,\ee^{-\ii \epsilon \tau^{2}}\right].
\end{align}
\par
When we recall the integral relation, \cref{eq:integral:relation}, we find 
\begin{align}
    \mathcal{I}_{s} = \frac{1}{2}\sqrt{\frac{\pi}{\epsilon}}\Im\left[\ee^{\ii \left(\beta-\pi/4\right)}\right].
\end{align}
Since the numerical integration, in complete agreement with the Markov result, \cref{eq:integral:over:varphi:dot:whole:time}, shows $\varphi (\infty) = 0$, we require the choice 
\begin{align}
    \beta = \frac{\pi}{4}
    \label{eq:beta:value}
\end{align}
for the constant $\beta$ of integration.

\subsubsection{Determination of amplitude}
The condition for $\beta$ does not specify the amplitude $\mathcal{S}$ of the oscillatory term in $\dot{\varphi}_{h}$. To find $\mathcal{S}$, we first simplify the formula, \cref{eq:general:solution:for:A}, for $A$, consistent with the approximations leading to $\dot{\varphi}_{h}$.
\par
For this purpose, we recall that the linear differential equation, \cref{eq:differential:equation:without:minus:four:term}, emerges from the non-linear differential equation, \cref{eq:differential:equation:for:phase}, when we neglect the non-linear terms. We then perform the integration and match the result to the well-known Landau-Zener formula. 
\par
We start from the amplitude
\begin{align}
    A(\infty) = \exp\left[-\frac{1}{2} \int_{-\infty}^{\infty} \dd \tau \frac{\ddot{\varphi}(\tau)}{\dot{\varphi}(\tau) + \epsilon \tau}\right].
    \label{eq:amplitude:A:at:infty}
\end{align}
in the asymptotic limit of infinite times, and given by \cref{eq:general:solution:for:A}.
\par
Next, we neglect the non-linearity, that is we drop the term $\dot{\varphi}$ in the denominator which leads us to the expression
\begin{align}
    \mathcal{A}(\infty) \equiv \exp\left[-\int_{-\infty}^{\infty} \dd \tau^{\prime} \frac{\ddot{\varphi}(2\tau^{\prime})}{\epsilon \tau^{\prime}}\right].
\end{align}
The \textit{anti-symmetric} part of the phase velocity leads to a \textit{symmetric} contribution to $\ddot{\varphi}$ which due to the term $\epsilon\tau$ in the denominator makes it into an anti-symmetric function. Hence, this part cancels out in this integral and only the oscillatory part $\dot{\varphi}_{h}$ for positive parts survives, leading us to the expression
\begin{align}
    A(\infty) \approx \exp\left[-\int_{0}^{\infty} \dd \tau \frac{\ddot{\varphi}_{h}(\tau)}{2\epsilon \tau}\right].
\end{align}
\par
With the help of \cref{eq:varphi:dot:dot:homogeneous:solution} the asymptotic $A(\infty)$ amplitude takes the explicit form
\begin{align}
    A(\infty) = \ee^{-\mathcal{S}\mathcal{I}_{c}}
\end{align}
where we have introduced the integral
\begin{align}
    \mathcal{I}_{c} \equiv \int_{0}^{\infty} \dd \tau \cos(\beta - \epsilon \tau^{2}) = \Re \left[\ee^{\ii \beta} \int_{0}^{\infty} \dd \tau\,\ee^{-\ii \epsilon \tau^{2}}\right].
\end{align}
\par
We again recall the integral relation, \cref{eq:integral:relation}, and find 
\begin{align}
    \mathcal{I}_{c} = \frac{1}{2}\sqrt{\frac{\pi}{\epsilon}}\Re\left[\ee^{\ii \left(\beta-\pi/4\right)}\right].
\end{align}
which with the expression, \cref{eq:beta:value}, for the phase shift $\beta$ reduces to
\begin{align}
    \mathcal{I}_{c} = \frac{1}{2}\sqrt{\frac{\pi}{\epsilon}}.
\end{align}
\par
When we match the product $\mathcal{S}\mathcal{I}_c$ with the exact asymptotic result $I$ given by \cref{eq:for:I}, we obtain the condition
\begin{align}
    \mathcal{S}\cdot\frac{1}{2}\sqrt{\frac{\pi}{\epsilon}} = \frac{\pi}{2\epsilon}
\end{align}
which leads us to the expression 
\begin{align}
    \mathcal{S} = \sqrt{\frac{\pi}{\epsilon}}
    \label{eq:S}
\end{align}
for the amplitude $\mathcal{S}$ of the oscillatory term.

\subsubsection{Summary}
This approach brings out most clearly that the final Landau-Zener probability amplitude is the result of three features:
(i) The oscillatory term of the phase velocity has the amplitude $\mathcal{S}$ given by \cref{eq:S}, varies quadratically in time, and has a phase shift of $\pi/4$. (ii) The probability amplitude is an integral over positive times only, and (iii) the integration yields another factor $\mathcal{S}/2$. 
\par
In the language of John A. Wheeler this interpretation takes the form
\begin{align}
\begin{pmatrix}\mathrm{Landau-}\\\mathrm{Zener}\\\mathrm{probability}\\\mathrm{amplitude}\end{pmatrix} = \exp\left[-\begin{pmatrix}\mathrm{amplitude}\\\mathrm{of}\\\mathrm{oscillations}\\\mathrm{in~phase~velocity}\end{pmatrix}\times\begin{pmatrix}\mathrm{integral}\\\mathrm{over}\\\mathrm{quadratic}\\\mathrm{phase}\end{pmatrix}\right]
\end{align}
that is 
\begin{align}
    a(\infty) = \exp\left[-\mathcal{S}\cdot\frac{1}{2}\mathcal{S}\right]
\end{align}
where the factor $1/2$ reflects the fact that the integration extends only over positive times.

\section{Conclusion and outlook}
\label{section:conclusion:outlook}
In the present article, we have revisited the well-known problem of Landau-Zener transitions. Here, we have first recalled the essential ingredients of our recent approach \cite{Glasbrenner2023} based on the Markov approximation. The corresponding integro-differential equation for the probability amplitude $a$ reduces to the elementary differential equation of an exponential function when we factor $a$ out of the integral. The essence of the Markov approximation is the dependence of the probability amplitude at time $\tau$ solely on $\tau$, and not on the values at previous times. This, at first sight, dramatic approximation provides us not only with the exact asymptotic Landau-Zener value, but also connects the behavior of the probability amplitude at large negative times with the one at large positive times with the help of Fresnel integrals. When we evaluate these integrals for large positive and negative times, we obtain approximate but analytic expressions for the probability amplitude. 
\par
We have complemented the Markov approach by one based on a decomposition of the probability amplitude into its amplitude and phase. The resulting equations of motion are coupled but lead, when decoupled, to rather complicated equations for only the amplitude and only the phase. However, when we neglect the non-linear terms, or consider the asymptotic limit of large negative and large positive times, we can solve these equations approximately and confirm the analytic expressions obtained by the Markov technique. 
\par
However, there is a subtlety. In contrast to the Markov solution, we now do not have an analytic expression that connects the two asymptotic time domains. We have to solve the differential equations in the two different regimes separately. This complication is not a problem for large negative times since we have the initial condition to fix the solution. However, it is a problem for large positive times. Here, the approach cannot determine the phase and the amplitude of the Stueckelberg oscillations. These constants can only be obtained by a comparison to the Markov approximation.
\par
Obviously, many questions remain. It suffices to name three.
\par
The amplitude-and-phase approach yields results identical to the Markov solution when we neglect the non-linearity. Hence, there must be a strong connection between linearization and Markov approximation, but what is this connection and why does the Markov solution provide us still with the exact result neglecting the non-linearity.
\par
So far, we have only focused on the probability amplitude $a$, and we may ask the following question: Do our techniques lead to meaningful results when applied to the probability amplitude $b$?
\par
Finally, a major complication of the Landau-Zener problem is the time dependence of the Hamiltonian, \cref{eq:time:dependent:hamiltonian}, requiring time-ordering of the corresponding Dyson series. However, in an unpublished article \cite{rojo2010} A. G. Rojo has demonstrated by an elegant evaluation of the high-dimensional  coupled integrals that the Dyson sum is just the exponential of the familiar Landau-Zener result. Unfortunately, Rojo did not point out the deeper symmetry, which made this calculation possible. Is it contained in the rapidly varying quadratic phase factors or in an attractor of the set of differential equations?
\par
Although we have gained many insights into and numerous answers to some aspects of these questions, it still represents work in progress. Therefore, we have to postpone the publication of these answers to the 150th birthday of Victor.

\begin{acknowledgments}
We thank M. Efremov, D. Fabian and A. Friedrich for many fruitful discussions.
W.P.S. is most grateful to Texas A\&M University for a Faculty Fellowship at the Hagler Institute for Advanced Study at Texas A\& M University and to Texas A\&M AgriLife for the support of this work. The research of the IQ\textsuperscript{ST} is financially supported by the Baden-Württemberg Ministry of Science, Research and Arts.
\end{acknowledgments}

\nocite{*}

\bibliography{references}% Produces the bibliography via BibTeX.

\end{document}